# Memory Window in Ferroelectric Field-Effect Transistors: Analytical Approach

Kasidit Toprasertpong, Member, IEEE, Mitsuru Takenaka, Member, IEEE, and Shinichi Takagi, Senior Member, IEEE

*Abstract*—A memory window of ferroelectric field-effect transistors (FeFETs), defined as a separation of the HIGH-state and the LOW-state threshold voltages, is an important measure of the FeFET memory characteristics. In this study, we theoretically investigate the relation between the FeFET memory window and the *P-E* hysteresis loop of the ferroelectric gate insulator, and derive a compact model explicitly described by material parameters. It is found that the memory window is linearly proportional to the ferroelectric polarization for the small polarization regime, and converges to the limit value of 2×coercive field× thickness when the remanent polarization is much larger than permittivity×coercive field. We discuss additional factors that possibly influence the memory window in actual devices such as the existence of interlayer (no direct impact), interface charges (invalidity of linear superposition between the ferroelectric and charge-trapping hysteresis), and minor-loop operation (behavior equivalent to the generation of interface charges).

## I. Introduction

FERROELECTRIC devices have been considered promising for low-power, high-speed, and high-density nonvolatile memory due to spontaneous polarization's remarkable feature [1, 2]. By applying an electric field larger than the coercive field, the polarization in the ferroelectric material can be switched to UP or DOWN states, which are stored as binary states in ferroelectric memory. A ferroelectric field-effect transistor (FeFET) is a form of ferroelectric memory that works with nondestructive readout operation [3]. It has been getting great attention in recent years after the discovery of HfO$_2$-based ferroelectric material owing to the CMOS compatibility [4], the satisfactory interface property with silicon [5], and the improved reliability [6,7]. A ferroelectric material is employed as a gate insulator in a field-effect transistor (FET) to control the conductivity of the semiconducting channel of an FeFET [8]. The ferroelectric polarization controls the FeFET threshold voltage $V_{th}$ and the stored polarization state is read out from the channel conduction.

A memory window (MW), defined as a separation of the HIGH threshold gate voltage $V_{th,H}$ at the UP-state polarization and the LOW threshold gate voltage $V_{th,L}$ at the DOWN-state polarization

$$\mathrm{MW} \equiv V_{th,H} - V_{th,L}, \qquad (1)$$

is a critical parameter and is considered one figure of merit of FeFETs as memory devices. The MW should be large enough to ensure a sufficient read margin and thus minimize the read error rate possibly caused by noise and disturb [9]. For this reason, it is essential to acquire a thorough understanding of how the MW of FeFETs is determined by the ferroelectric properties of the ferroelectric insulator such as the remanent polarization $P_r$, the coercive field $E_c$, and the dielectric constant $\varepsilon_{FE}$ as well as possible dependency on other factors. The device understanding would help us take the proper design direction and realize FeFETs with satisfactory properties for nonvolatile memory.

Despite the need for the understanding of the behavior of MW in the FeFET operation, there is a lack of a closed-form analytical expression that can comprehensively describe the characteristics of MW. S. L. Miller and P. J. McWhorter [8] have summarized a set of the governing equations of FeFETs, but the behavior of MW has not been sufficiently discussed. Most research works discussing about MW so far have been studied through self-consistent simulations solving a set of governing equations [10-13]; however, such a numerical approach provides limited information since the role of each material parameter as well as the underlying physical mechanism is difficult to interpret accurately. One approximate expression MW = $2E_c t_{FE}(1-E_c\varepsilon_{FE}\varepsilon_0/\eta P_s)$ ($P_s$: spontaneous polarization; $\eta$: squareness of hysteresis loop; $t_{FE}$: ferroelectric thickness; $\varepsilon_0$: vacuum permittivity) has been proposed by Lue *et al.* [14], but this expression only holds at large $P_s$ otherwise the expression will be unrealistically negative. Even though a model based on the Landau-Devonshire theory has also been proposed [15], the theory assumes ideal ferroelectric with a single polarization domain and a more generalized model is needed to describe the actual behavior of practical FeFETs.

In this paper, we investigate the mechanism governing the behavior of MW and propose an analytical closed-form expression for the MW of FeFETs. The analytical expression

This work was supported by JSPS KAKENHI Grant No. 21H01359 and JST-CREST Grant No. JPMJCR20C3, Japan.

The authors are with the Department of Electrical Engineering and Information Systems, The University of Tokyo, Tokyo 113-8656, Japan (e-mail: toprasertpong@mosfet.t.u-tokyo.ac.jp; takenaka@mosfet.t.u-tokyo.ac.jp; takagi@ee.t.u-tokyo.ac.jp).

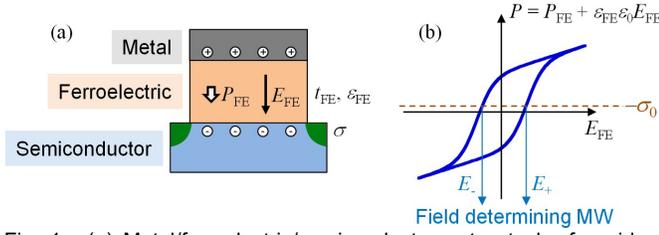

Fig. 1. (a) Metal/ferroelectric/semiconductor gate stack of an ideal FeFET. (b) Graphical approach to determine the operating field at the threshold voltages from the *P-E* characteristics. The operating fields are determined by the intercept points between the *P-E* characteristics and the horizontal line of $P = -\sigma_0$.

allows us to directly examine the impacts of the ferroelectric parameters on the MW and discuss the appropriate ferroelectric parameters to improve the MW. Possible impacts of nonideal parameters in FeFETs such as nonferroelectric interlayers, trapped and fixed charges, and minor-loop on MW are discussed.

## II. MEMORY WINDOW IN IDEAL FEFETS

We first consider the MW of ideal FeFETs whose gate stack is shown in Fig. 1(a). Even though ideal FeFETs are not easily achieved in real devices, it is necessary to study the intrinsic properties of FeFETs to understand their theoretical limit and properly assess the impact of extrinsic factors. Ideal FeFETs here refer to those with metal/ferroelectric/semiconductor (MFS) gate stacks that include no interlayer and no interface charges, and the ferroelectric polarization operates in the saturated hysteresis loop. Unless otherwise specified, parameters of ferroelectric HfO$_2$ ($E_c$ = 1.5 MV/cm, $\varepsilon_{FE}$ = 30, $t_{FE}$ = 10 nm [5,16]) are used in the calculation, but the same conclusion is still valid for other ferroelectric materials.

As stated in [8,14], the relation between the ferroelectric polarization $P_{FE}$, electric field $E_{FE}$ in the ferroelectric layer, and the semiconductor charge density $\sigma$ is given by Gauss' law:

$$P_{FE} + \varepsilon_{FE}\varepsilon_0 E_{FE} = -\sigma . \quad (2)$$

The minus sign on the right comes from the directions of $P_{FE}$ and $E_{FE}$ defined in Fig. 1(a). $E_{FE}$ here is the total electric field that includes the influences from the gate voltage $V_g$, the depolarization field, and possible trapped charges (see Sec. IIIB). The threshold voltage $V_{th}$ of FET devices is the gate voltage $V_g$ at which the semiconductor channel becomes conductive. We can generally consider that FETs, independent of the operation modes (e.g., inversion or junctionless), are at the threshold when the semiconductor reaches a certain surface potential $\psi_s$ and the semiconductor charge density $\sigma$ reaches a threshold value $\sigma_0$. This suggests that the polarization states in FeFETs at the threshold voltage must satisfy

$$P_{FE\pm}(E_\pm) + \varepsilon_{FE}\varepsilon_0 E_\pm = -\sigma_0 \quad \text{at } V_g = V_{th,H} \text{ or } V_{th,H}, \quad (3)$$

where $P_{FE\pm}$ is the ferroelectric polarization and $E_\pm$ is the electric field in the ferroelectric layer at $V_{th,H}$ (+) and $V_{th,L}$ (−), respectively.

The gate voltage $V_g$ of FeFET is given by the electric field $E_{FE}$, the metal-semiconductor work function difference $\phi_{ms}$, and the surface potentials $\psi_s$ through $V_g = E_{FE}t_{FE} + \psi_s + \phi_{ms}$ [8,14]. As $\phi_{ms}$ is a constant and $\psi_s$ in these two states are the same according to the definition of threshold voltage, the MW is then determined by

$$\begin{aligned} MW &= V_{th,H} - V_{th,L} \\ &= (E_+ t_{FE} + \psi_s + \phi_{ms}) - (E_- t_{FE} + \psi_s + \phi_{ms}) \\ &= (E_+ - E_-)t_{FE} . \quad (4)\end{aligned}$$

Equation (4) implies that determining the operating electric fields $E_\pm$ at the threshold voltage is the key to estimating the FeFET MW. As the left side of (3) is actually the polarization hysteresis loop obtained by typical *P-E* measurement from a single ferroelectric layer (not from an MFS capacitor or from capacitors with more layers as $E_{FE}$ cannot be accurately measured in that case), we can obtain the operating electric field by a graphical approach as shown in Fig. 1(b). That is, the MW of ideal FeFETs is determined by the intercept points between the *P-E* loop and the horizontal line of $P = -\sigma_0$.

To discuss the impact of material properties, we employ the tanh model,

$$P_{FE\pm} = P_s \tanh(\eta(E_{FE} \mp E_c)/E_c); \quad \tanh(\eta) = P_r/P_s . \quad (5)$$

Equation (5) is a well-accepted expression to describe the steady-state characteristics of a wide range of ferroelectric materials [8,17,18]. Note that (5) is an empirical expression and does not consider the physical mechanisms inside specific ferroelectric materials. The model can be extended by including the mechanisms such as polarization reversal dynamics, which are material-specific and would be beyond the scope of this study.

Furthermore, considering that the polarization in ferroelectric materials is typically in the order of 1 to 100 μC/cm$^2$ whereas the threshold charge density at the threshold voltage is usually in the order of $|\sigma_0|$ = 10$^{-5}$ to 10$^{-2}$ μC/cm$^2$ (10$^8$ to 10$^{11}$ cm$^{-2}$) we may consider the right side of (3) to be negligibly small. From (3) and (5), the equation governing the MW behavior in ideal FeFETs is expressed by

$$P_s \tanh(\eta(E_\pm \mp E_c)/E_c) + \varepsilon_{FE}\varepsilon_0 E_\pm = 0 . \quad (6)$$

Figs. 2(a)-(c) shows the MW as a function of $P_r$ (=$P_s\tanh(\eta)$) simulated by simultaneously solving the above relations with various values of $E_c$, $\varepsilon_{FE}$, and $P_r/P_s$. It can be seen that there are two important operating regimes: the small $P_r$ regime where MW is linearly proportional to $P_r$, and the large $P_r$ regime where MW almost converges to a particular value with a strong dependency on $E_c$ but not on $P_r$.

To better understand this behavior, we consider approximating (6) by closed-form expressions. In the case where $P_r \ll \varepsilon_{EF}\varepsilon_0 E_c$, the first term becomes small and the absolute values of $E_\pm$ in the second term have to be small to balance both terms. Therefore, we may neglect $E_\pm$ in the tanh function of the first term and rewrite (6) by

$$E_\pm \approx \frac{P_s}{\varepsilon_{FE}\varepsilon_0}\tanh(\pm\eta) = \pm\frac{P_r}{\varepsilon_{FE}\varepsilon_0} \text{ when } P_r \ll \varepsilon_{FE}\varepsilon_0 E_c. \quad (7)$$

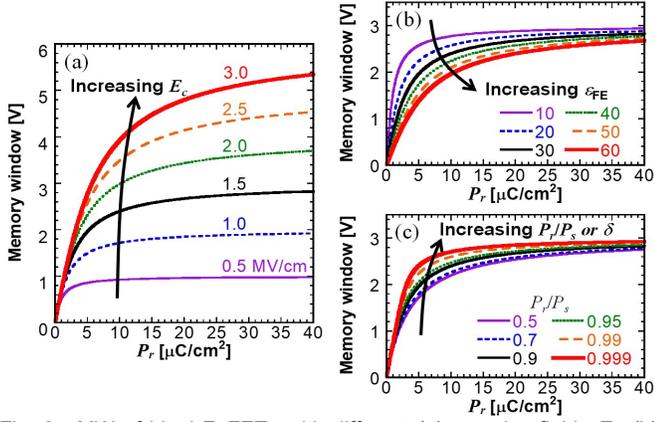

Fig. 2. MW of ideal FeFETs with different (a) coercive fields $E_c$, (b) dielectric constants $\varepsilon_{FE}$, and (c) $P_r/P_s$ ratios or tanh($\eta$). When values are not specified, $E_c$ = 1.5 MV/cm, $\varepsilon_{FE}$ = 30, $P_r/P_s$ = 0.9, and $t_{FE}$ = 10 nm are used as default parameters in the simulations.

On the other hand, at the large $P_r$ regime, the value in the tanh function in (6) has to be small to balance the equation. Since $\tanh(x) \approx x$ holds when $|x| < 0.5$ (c.f. $\tanh(0.5) \approx 0.46$)), (6) in the large $P_r$ regime can be rewritten by

$$P_s\left(\eta(E\mp E_c)/E_c\right) + \varepsilon_{FE}\varepsilon_0 E_\pm \approx 0$$
$$\Leftrightarrow E_\pm \approx \pm\frac{E_c}{1+\varepsilon_{FE}\varepsilon_0 E_c/\eta P_s} = \pm\frac{E_c}{1+\varepsilon_{FE}\varepsilon_0 E_c/P_r \cdot (\tanh\eta/\eta)}$$
$$\text{when } P_r > \varepsilon_{FE}\varepsilon_0 E_c \tanh(\eta)(2-1/\eta). \quad (8)$$

The valid range in (8) is estimated by substituting $\eta|E_\pm \mp E_c|/E_c < 0.5$, which is the condition to ensure $\tanh(\eta|E_\pm \mp E_c|/E_c) \approx \eta|E_\pm \mp E_c|/E_c$, back into (8). From (4), (7), (8), the MW can be written by the following closed-form expressions

$$\text{MW} \approx \begin{cases} \dfrac{2P_r t_{FE}}{\varepsilon_{FE}\varepsilon_0} & ; P_r \ll \varepsilon_{FE}\varepsilon_0 E_c \quad (9a) \\[1em] \dfrac{2E_c t_{FE}}{1+\dfrac{\varepsilon_{FE}\varepsilon_0 E_c}{P_r}\dfrac{\tanh\eta}{\eta}} & ; P_r > \varepsilon_{FE}\varepsilon_0 E_c\tanh(\eta)(2-\dfrac{1}{\eta}). \quad (9b) \end{cases}$$

Figs. 3(a)-(d) show the results obtained from the approximate expressions (9) in dashed lines and the exact solutions solving with (3)-(6) in solid lines. We can confirm that the small $P_r$ and large $P_r$ regimes can be well explained by 9(a) and 9(b), respectively, validating the approximate expressions (9). Fig. 3(a) also shows that for $\eta \leq 0.5$ ($P_r/P_s < 0.5$), the expression 9(b) alone is a good approximation for the entire $P_r$ range as can

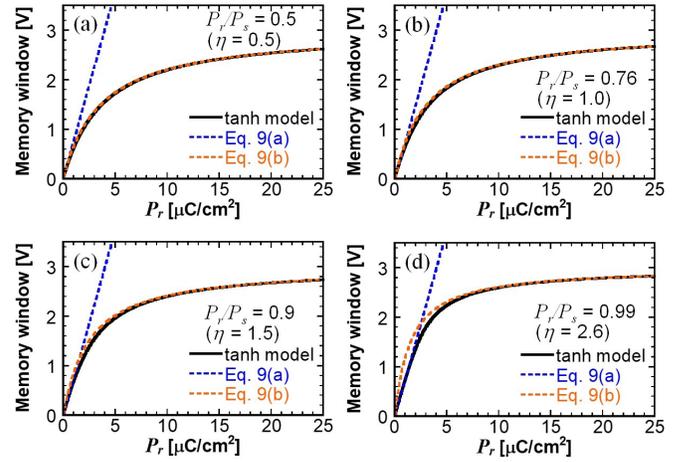

Fig. 3. Approximation of MW by (9) for of FeFETs with $P_r/P_s$ = (a) 0.5, (b) 0.76, (c) 0.9, and (d) 0.99 ($E_c$ = 1.5 MV/cm, $\varepsilon_{FE}$ = 30, and $t_{FE}$ = 10 nm). MWs at small $P_r$ can be approximated by 9(a), whereas those at large $P_r$ can be well approximated by 9(b). For low $P_r/P_s$ (low $\eta$), only 9(b) is enough to approximate the entire range of $P_r$.

be expected from the valid range of (9b). In this case, (9b) converges to (9a) at small $P_r$. Note that at the limit of large $P_r$, (9b) converges to the same limit as $2E_c t_{FE}(1-E_c\varepsilon_{FE}/\eta P_s)$ proposed in [14].

Since (9a) and (9b) are simple functions of materials parameters, it is straightforward to examine how the MW is determined. One interesting feature is that as long as the threshold charge density $\sigma_0$ is much smaller than the ferroelectric polarization, the MW of ideal FeFETs is determined only by the material parameters of the ferroelectric insulator and is independent of the types of metal and semiconductor. At the small $P_r$ regime, the MW is linearly proportional to $P_r$ while independent of $E_c$. At the large $P_r$ regime, the roles of $E_c$ and $P_r$ become opposite: the MW converges to its maximum value of $2E_c t_{FE}$ and is only weakly dependent on $P_r$. We can understand the behavior of the large and small $P_r$ regimes in this way. When the ferroelectric polarization is the main polarization component, the FeFET state is controlled by the ferroelectric polarization reversal and thus the coercive field $E_c$. On the other hand, when the ferroelectric component is smaller than the linear component $\varepsilon_{EF}\varepsilon_0 E$, the sign of total polarization can be modulated even without the ferroelectric polarization reversal. The sign reversal occurs when the linear component cancels out the ferroelectric component, $\varepsilon_{EF}\varepsilon_0 E = P_r$, corresponding to (7) and (9a).

From these trends, it is suggested that $P_r$ needs to be improved when $P_r$ is still small, but a drastic improvement is not always necessary as only a small improvement of MW can be expected at very large $P_r$. It is important to discuss here what should be the design target of $P_r$ for the FeFET application. Let's consider the large $P_r$ regime in (9b). If we aim for the 75% of the maximum MW of $2E_c t_{FE}$, we need $P_r$ of at least

$$P_{r,>75\%} \approx 3\varepsilon_{FE}\varepsilon_0 E_c. \quad (10)$$

$P_{r,>75\%} = 3\varepsilon_{FE}\varepsilon_0 E_c \tanh(\eta)/\eta$ may be used instead if $\eta$ is given. Figs. 4(a),(b) show the MW normalized to $2E_c t_{FE}$ for ferroelectric HfO$_2$ and SrBi$_2$Ta$_2$O$_9$ (SBT) using the parameters from [5,16]. Pb(Zr$_x$Ti$_{1-x}$)O$_3$ (PZT) has coercive field and dielectric constant depending on the Zr composition [19], but Figs. 4(a),(b) can also be applied to PZT as well as other ferroelectric materials if the $E_c \times \varepsilon_{FE}$ product falls into the similar range. For ferroelectric HfO$_2$ ($E_c$ = 1.5 MV/cm and $\varepsilon_{FE}$ = 30), $\varepsilon_{FE}\varepsilon_0 E_c$ = 4 μC/cm$^2$ and thus $P_r$ = 7-12 μC/cm$^2$ ($\eta$ = 0-1.5), or roughly 10 μC/cm$^2$, is necessary to obtain a large MW. For ferroelectric SBT ($E_c$ = 0.05 MV/cm and $\varepsilon_{FE}$ = 200), the $\varepsilon_{FE}\varepsilon_0 E_c$ is 0.9 μC/cm$^2$ and thus only small $P_r$ = 1-3 μC/cm$^2$ is necessary to make the MW value close to the maximum.

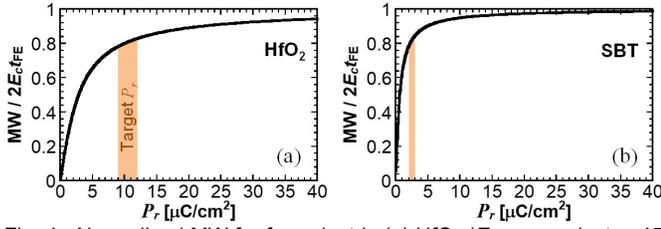

Fig. 4. Normalized MW for ferroelectric (a) HfO$_2$ ($E_c \times \varepsilon_{FE}$ product ≈ 45 MV/cm) and (b) SBT ($E_c \times \varepsilon_{FE}$ product ≈ 10 MV/cm). $P_r/P_s$ = 0.9 is assumed. The shaded areas indicate the target value of $P_r \sim 3\varepsilon_{FE}\varepsilon_0 E_c$ to obtain large MW. $P_r$ larger than this target is also possible, but only a small improvement of MW is expected.

It should be noted that the suitable $P_r$ may be deviated from (10) if the endurance and retention issues in practical devices have to be considered. Too large $P_r$ may result in a poor endurance property as it induces enormous interface charges which degrades FET interface [10,12,20,21]. In this case, smaller $P_r$ with some MW penalty can be preferable to mitigate the endurance degradation. For instance, we need only $P_r \approx \varepsilon_{FE}\varepsilon_0 E_c$ if we aim for MW = 0.5×2$E_c t_{FE}$, indicating that largely decreasing $P_r$ results in a small MW penalty. The retention property is also another concern in practical devices since $P_r$ induces a depolarization field which degrades the retention time. The optimum $P_r$ value considering the reliability issue depends on materials, structures, and target applications. We believe that the analytical expression of MW in (9) would be helpful when the simultaneous optimization of these multiple factors is needed.

If high $P_r$ can be achieved, MW is in the order of $E_c t_{FE}$ and can be improved by using ferroelectric materials with large $E_c$ (Fig. 2(a)). Moreover, 9(a) and 9(b) suggest that large $\varepsilon_{FE}$ has a negative impact on the MW of FeFETs for the entire $P_r$ range as demonstrated in Fig. 2(b). Therefore, it is suggested that (i) $P_r$ of the order of $\varepsilon_{FE}\varepsilon_0 E_c$, (ii) small $\varepsilon_{FE}$, and (iii) large $E_c$ are required to obtain a large memory window, while these design targets, particularly (i), can be deviated depending on the device reliability requirement for each application. Even though SBT easily satisfies (i), it has high $\varepsilon_{FE}$ and small $E_c$. Particularly the small $E_c$ causes SBT to require thickness $t_{FE}$ of more than 100 nm to guarantee MW (< 2$E_c t_{FE}$) in the order of 1 V for the memory application.

### III. NONIDEAL FACTORS IN PRACTICAL DEVICES

We will discuss additional factors that can possibly affect the MW characteristics of nonideal FeFETs.

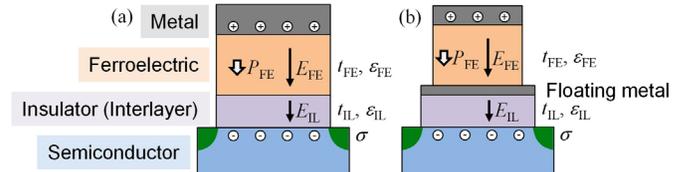

Fig. 5. Schematic of (a) MFIS-type and (b) MFMIS-type FeFETs.

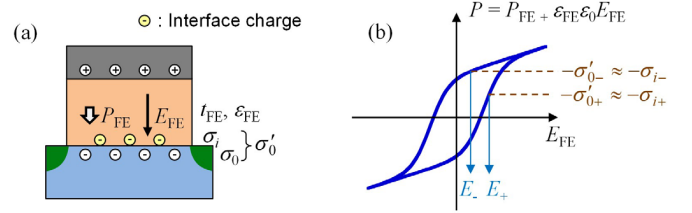

Fig. 6. (a) Schematic and (b) operating points of FeFET with interface charges $\sigma_i$.

#### A. MFIS-type and MFMIS-type FeFETs

The schematic of an FeFET with an MFIS-type (M: Metal; F: Ferroelectric; I: Insulator or Interlayer; S: Semiconductor) structure is shown in Fig. 5(a). In this structure, Gauss' law at the threshold voltages ($\sigma = \sigma_0$) is

$$P_{FE} + \varepsilon_{FE}\varepsilon_0 E_{FE} = \varepsilon_{IL}\varepsilon_0 E_{IL} = -\sigma_0 . \quad (11)$$

Since the additional voltage across the gate stack due to the interlayer is equally $E_{IL} t_{IL} = -\sigma_0 t_{IL}/\varepsilon_{IL}\varepsilon_0$ (from (11)) at both the LOW and HIGH threshold voltages, the MW is determined by

$$\begin{aligned} MW &= (E_+ t_{FE} - \sigma_0 t_{IL}/\varepsilon_{IL}\varepsilon_0 + \psi_s + \phi_{ms}) \\ &\quad - (E_- t_{FE} - \sigma_0 t_{IL}/\varepsilon_{IL}\varepsilon_0 + \psi_s + \phi_{ms}) \\ &= (E_+ - E_-) t_{FE} . \end{aligned} \quad (12)$$

Note that $E_{FE}$ and $E_\pm$ are the total electric field that already includes the impact of the depolarization field which might be enhanced by the existence of the interlayer. Since (11) and (12) are the same as (2) and (4), all discussions so far in Sec. II is still valid and we can say that the interlayer in MFIS-FeFETs has no direct effect on the MW as far as the ferroelectric layer operates in the saturated hysteresis loop. Note that if write voltage is limited, the existence of an interlayer decreases the voltage across the ferroelectric layer and may affect the MW through minor-loop operation (see Sec. IIIC).

For MFMIS-type FeFETs (Fig. 5(b)), the area ratio of the ferroelectric to the insulator ($A_{FE}/A_{IL}$) is an additional structure parameter:

$$P_{FE} + \varepsilon_{FE}\varepsilon_0 E_{FE} = \varepsilon_{IL}\varepsilon_0 E_{IL}(A_{IL}/A_{FE}) = -\sigma_0(A_{IL}/A_{FE}) . \quad (13)$$

This indicates that the discussion so far is still valid. However, when $A_{FE}/A_{IL}$ is extremely small (e.g., less than 1/100), the effect of the right side of (13) has to be considered, which is equivalent to interface charges discussed in Sec. IIIB.

#### B. Interface charges

We consider the effect of additional charges $\sigma_i$ at the interface that does not contribute to the FeFET channel current as shown in Fig. 6(a). First, an MFS interface and only charges at the ferroelectric/semiconductor interface are considered for simplicity. Gauss' law at the threshold voltages ($\sigma = \sigma_0$) has to be rewritten to

$$P_{FE} + \varepsilon_{FE}\varepsilon_0 E_{FE} = -\sigma_0' \approx -\sigma_i, \quad (14)$$

where $\sigma_0' = \sigma_0 + \sigma_i$ is the total threshold charges considering both the threshold semiconductor charges $\sigma_0$ and the interface charges $\sigma_i$. In most cases where $\sigma_0$ is small, $\sigma_0' \approx \sigma_i$ can be used instead as shown on the right of (14). We denote $\sigma_{i+}$ and $\sigma_{i-}$ as the interface charges at the HIGH and LOW threshold voltages, respectively. The MW is determined by MW = ($E_+$ - $E_-$)$t_{FE}$ and the operating fields $E_\pm$ can be determined by the graphical approach considering the cross points with the $P$-$E$ hysteresis loop as shown in Fig. 6(b). In the case of MFIS-FeFETs (and similarly MFMIS-FeFETs), the interface charges at the ferroelectric/interlayer interface modify Gauss' law to [22]

$$P_{FE} + \varepsilon_{FE}\varepsilon_0 E_{FE} + \sigma_i = \varepsilon_{IL}\varepsilon_0 E_{IL} = -\sigma_0, \quad (15)$$

which indicates that $E_{IL}t_{IL} = -\sigma_0 t_{IL}/\varepsilon_{IL}\varepsilon_0$, (12), and (14) are still valid. In other words, the following discussion is also applicable for MFIS- and MFMIS-FeFETs with additional charges at the ferroelectric/interlayer interface.

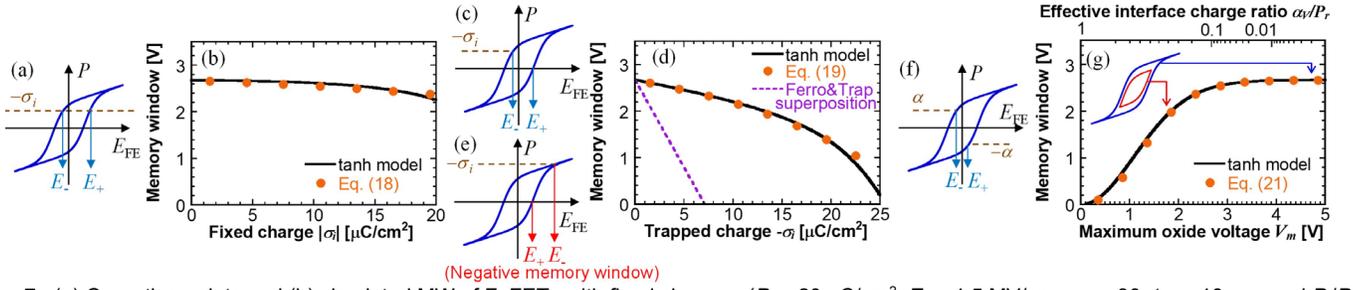

Fig. 7. (a) Operating points and (b) simulated MW of FeFETs with fixed charges. ($P_r$ = 20 μC/cm$^2$, $E_c$ = 1.5 MV/cm, $\varepsilon_{FE}$ = 30, $t_{FE}$ = 10 nm, and $P_r/P_s$ = 0.9) (c),(d) FeFETs with trapped negative charges after positive write operation. A dashed line shows the MW estimated from linear superposition between the ferroelectric MW (MW$_0$) and the charge trapping hysteresis (-$\sigma_i t_{FE}/\varepsilon_{FE}\varepsilon_0$). (e) Operating points when the negative MW is induced by large trapped charges. (f),(g) FeFETs under minor-loop operation with limited write voltage. The minor-loop operation has a similar behavior to the saturation-loop operation with the effective interface charge $\alpha_V$.

To describe the impact of the interface charges on the MW by an analytic expression, we need to solve the operating fields $E_\pm$ from

$$P_\pm(E_\pm) = P_s \tanh(\eta(E_\pm \mp E_c)/E_c) + \varepsilon_{FE}\varepsilon_0 E_\pm = -\sigma_{i\pm}, \quad (16)$$

Unfortunately, an analytic expression of $E_\pm$ from (16) is not easily obtained for large $\sigma_{i\pm}$. To deal with this issue, let's consider an approximated inverse function of (16),

$$E_\pm(P) \approx P/\varepsilon_{FE}\varepsilon_0 + E_{\pm,P0} - \xi_\pm \tanh(\theta_\pm P/P_s); \; P \geq 0, \quad (17a)$$

$$\theta_\pm \approx \tfrac{1}{2}(1 \mp \tfrac{3}{5}\varepsilon_{FE}\varepsilon_0 E_c/P_s\eta), \quad (17b)$$

$$\xi_- = (P_r/\varepsilon_{FE}\varepsilon_0 + E_{-,P0})/\tanh(\theta_- P_r/P_s), \; \xi_+ = \xi_- + 2E_{+,P0}, \quad (17c)$$

$$E_\pm(-P) = -E_\mp(P). \quad (17d)$$

Eq. (17) is a fitting model that is confirmed to be a good approximation for ferroelectric with $\varepsilon_{FE}\varepsilon_0 E_c/P_r = 0.1\sim0.5$. The coefficients $\theta_\pm$ and $\xi_\pm$ can be slightly adjusted for other ranges of $\varepsilon_{FE}\varepsilon_0 E_c/P_r$. It exhibits essential features of the hysteresis loop for a rough discussion on the impact of interface charges on the FeFET operation ($E_+ = E_{\pm,P0}$ when $P = 0$ ($E_{\pm,P0}$ are given by (7), (8)), $E_\pm \to P/\varepsilon_{FE}\varepsilon_0$ and $E_+ - E_- \to 0$ when $|P| \to \infty$, and $E_- = 0$ when $P = P_r$).

When the interface charges are fixed charges or trapped charges that are similarly induced at the HIGH and LOW threshold voltages, $\sigma_{i+} = \sigma_{i-} = \sigma_i$, the operating fields are determined by the cross points with the horizontal line $P = -\sigma_i$ as shown in Fig. 7(a). An approximate expression describing the impact of fixed charges on the MW is obtained from MW = ($E_+$ - $E_-$)$t_{FE}$ using (17) when $P = -\sigma_i$:

$$MW_{fixed\_charge} \approx MW_0 \left(1 - \tanh(\theta_+|\sigma_i|/P_s)\right)$$
$$- \xi_- t_{FE}\left(\tanh(\theta_+|\sigma_i|/P_s) - \tanh(\theta_-|\sigma_i|/P_s)\right), \quad (18)$$

where MW$_0$ is given by (9). This relation implies that the MW tends to be smaller for larger $\sigma_i$, confirmed from the narrowing field window at large $\sigma_i$ in Fig. 7(a) and the simulation results in Fig. 7(b). This suggests that if charges are injected into the ferroelectric interface during the FeFET operation and cannot be detrapped, the MW will decrease as the FeFET operation progresses.

Another possible situation is when trapped charges are asymmetric: $\sigma_{i+} = 0$ and $\sigma_{i-} < 0$. This situation can be seen when the electron trapping is induced at the ferroelectric/interlayer of MFIS-FeFETs after applying positive write voltage [20,21]. The schematic of the operating points is shown in Fig. 7(c) and the MW = ($E_+$ - $E_-$)$t_{FE}$ in this case is given by

$$MW_{trapped\_charge} \approx MW_0 - \frac{-\sigma_i t_{FE}}{\varepsilon_{FE}\varepsilon_0} + \xi_- t_{FE} \tanh\left(\theta_- \frac{-\sigma_i}{P_s}\right). \quad (19)$$

The existence of the third term indicates that the MW of FeFETs with trapped charges is *not* a simple linear superposition of the ferroelectric MW when there are no traps ($MW_0$) and the threshold shift by charge trapping when there is no ferroelectricity ($-\sigma_i t_{FE}/\varepsilon_{FE}\varepsilon_0$), as confirmed by simulation in Fig. 7(d). This is due to the nonlinear interaction between polarization, trapped charges, and electric field originating from the nonlinearity of *P-E* characteristics. Therefore, trapped charge density cannot be simply estimated from $\Delta V_{th}\varepsilon_{FE}\varepsilon_0/t_{FE}$ in a similar way as in nonferroelectric FETs.

Fig. 7(e) shows that the large density of asymmetric charge trapping may cancel out the ferroelectric MW or even result in a negative MW. The ferroelectric MW disappears when the trapped charge density is $P_-(E_+)$, or slightly over $P_r$ ($\neq$ $MW \times \varepsilon_{FE}\varepsilon_0/t_{FE}$). This situation can be seen when the charge density more than $P_r$ is trapped at the interface after positive write voltage but cannot be detrapped fast enough before the next readout operation [23].

### C. Limited write voltage and minor-loop operation

The analysis so far has assumed a sufficiently large write voltage by which the ferroelectric polarization operates under a saturated hysteresis loop. In practical operation, the write voltage is finite and minor-loop operation has to be taken into account. To analytically examine the impact of limited write voltage, we consider a simplified situation in which the maximum fields across the ferroelectric layer are the same for both the positive and negative write operations. When the field is symmetric, we can use the minor-loop model proposed in [14], which states that the polarization is approximated by

$$\begin{cases} P_{minor} = P_s \tanh\left(\eta(E \mp E_c)/E_c\right) + \varepsilon_{FE}\varepsilon_0 E_\pm \pm \alpha_V & (20a) \\ \alpha_V = \frac{1}{2}P_s\left[\tanh\left(\eta(V_m/t_{FE} + E_c)/E_c\right) \\ \qquad - \tanh\left(\eta(V_m/t_{FE} - E_c)/E_c\right)\right] & (20b) \end{cases}$$

Here, $V_m/t_{FE}$ and $V_m$ are the maximum electric field and voltage across the ferroelectric layer. $\alpha_V$ is a parameter between 0 and $P_r$ that represents a minor-loop operation and ensures a closed loop of polarization hysteresis at $|E| = V_m/t_{FE}$. An example of a minor-loop modeled by (20) is shown in the inset of Fig. 7(g). Then, the MW is determined by putting (20a) to $-\sigma_0 \approx 0$, which is in fact similar to (16) with $\sigma_{i\pm}$ replaced by $\pm\alpha_V$. In other words, the minor-loop operation has similar behavior to the effective interface charges of $+\alpha_V$ during the forward voltage sweep and $-\alpha$ during the backward sweep, as shown in Fig. 7(f). By substituting $\sigma_{i\pm} = \pm\alpha_V$ in (17), we obtain

$$MW_{minor} \approx MW_0 \left(1 - \frac{\tanh(\theta_-\alpha_V/P_s)}{\tanh(\theta_-P_r/P_s)}\right) \\ - \frac{2t\alpha_V}{\varepsilon_{FE}\varepsilon_0}\left(1 - \frac{P_r}{\alpha_V}\frac{\tanh(\theta_-\alpha_V/P_s)}{\tanh(\theta_-P_r/P_s)}\right). \quad (21)$$

At sufficiently large voltage $V_m$ ($\alpha_V = 0$), $MW_{minor}$ converges to $MW_0$ obtained so far in (9). $MW_{minor}$ becomes smaller for small $V_m$ and drops to 0 when $V_m = 0$ ($\alpha_V = P_r$), as shown in Fig. 7(g).

### D. Other nonideal factors

In actual FeFETs, there could be other nonideal factors that are not included in the discussions here. If the *P-E* characteristics including nonideal factors are given, we can extract the MW using similar approaches shown in Fig. 1(b) and Fig. 6(b). That is, the MW is given by $(E_+ - E_-)t_{FE}$ (see (12)), where $E_\pm$ is determined by the cross points between the given *P-E* characteristics with the $P = -\sigma_0$ line (Fig. 1(b)) if interface charges do not exist, or with the $P \approx \sigma_{i\pm}$ line (Fig. 6(b)) if interface charges exist. Moreover, if the *P-E* characteristics can be written in an analytical form and the cross points can be solved analytically, we can obtain the closed-form expression of the MW in a similar way to those described in (1)-(21). We should note that when the *P-E* relation becomes too complicated, particularly when an FeFET operate under an asymmetric minor-loop, the current forms of our MW formulae cannot be applied. In such a case, we need to apply a more sophisticated approximation to formulate the MW expression, or we can estimate the MW numerically from the cross points as described above.

In addition, it should be addressed that the model in this study is based on the assumption of the steady-state operation so that the ferroelectric polarization $P_{FE}$ is uniquely determined by the electric field $E_{FE}$ in the film through (5). In actual operations where the operating time is finite, there could be transient dynamics caused by the finite polarization switching speed and the transient coupling of polarization with free/trapped charges, which causes $P_{FE}$ to be a function of time and electric field history in addition to the applying field $E_{FE}$. In such a case, the memory window expression will deviate from that given in this work, and we need a numerical simulation to estimate the memory window for each specific case.

## IV. CONCLUSION

We have proposed the compact model for the memory window of FeFETs which directly describes the impact of each ferroelectric parameter and helps understand how to improve the FeFET memory window. The model suggests that the memory window is determined by the coercive field $E_c$ only when the remanent polarization $P_r$ is large, and becomes strongly dependent on $P_r$ if $P_r$ is small. A low $\varepsilon_{FE}$ and a high $E_c$ are found to be favorable for obtaining a large memory window. Considering the impact of interface charges and minor-loop operation as a shift of the operating point on the *P-E* loop allows us to analytically examine and gain an understanding of the behavior of those nonideal factors.